# Correlation length from cluster-diameter distribution

Wolfhard Janke and Stefan Kappler[a]*

[a]Institut für Physik, Johannes Gutenberg-Universität Mainz, 55099 Mainz, Germany

We report numerical estimates of correlation lengths in 2D Potts models from the asymptotic decay of the cluster-diameter distribution. Using this observable we are able to verify theoretical predictions for the correlation length in the disordered phase at the transition point for $q = 10$, $15$, and $20$ with an accuracy of about $1\% - 2\%$. This is a considerable improvement over previous measurements using the standard (projected) two-point function.

## 1. INTRODUCTION

The two-dimensional $q$-state Potts model [1] is one of the most thoroughly studied models of statistical physics. Since many quantities of interest are known exactly [2] it has repeatedly been used as a testing ground for more general ideas on, e.g., finite-size scaling (FSS) [3], the shape of energy or magnetization distributions [4] or partition function zeros [5]. A few years ago also an analytical formula for the correlation length $\xi_d(\beta_t)$ in the disordered phase at the first-order transition point $\beta_t$ for $q \geq 5$ has been added to the list of exactly known quantities [6,7]. For the correlation length $\xi_o(\beta_t)$ in the ordered phase, however, only heuristic arguments were available until very recently [8].

This motivated us to perform a numerical study with the goal to clarify conflicting predictions for the ratio $\xi_o/\xi_d$ at $\beta_t$ [9]. To test the employed numerical techniques, we started this project by first studying the disordered phase [10]. Using cluster algorithms for the update and standard (projected) two-point correlation functions for extracting the correlation length we could confirm the analytical formula only with an accuracy of about $10\% - 20\%$, depending on the value of $q$. The deviation is clearly of systematic nature and could be traced back to the unexpected importance of higher order excitations. Even though we worked on large lattices of linear size up to $\approx 28\xi_d$ and performed quite a large number of about $50\,000 - 100\,000$ statistically independent measurements, fits with sufficiently many correction terms turned out to be too unstable.

Here we report a reinvestigation of this problem by using a completely different estimator for the correlation length, the asymptotic decay of the cluster-diameter distribution as discussed in recent analytical work [8]. Numerically this observable turns out to be very well suited in the disordered phase, allowing for the first time a confirmation of the analytical formula for $\xi_d(\beta_t)$ with an accuracy of about $1\% - 2\%$.

## 2. SIMULATION

We simulated the standard Potts model partition function

$$Z = \sum_{\{s_i\}} e^{-\beta E}; \ E = -\sum_{\langle ij \rangle} \delta_{s_i s_j}; \ s_i = 1, \ldots, q, \quad (1)$$

where $i$ denote the lattice sites, $\langle ij \rangle$ are nearest-neighbor pairs, and $\delta_{s_i s_j}$ is the Kronecker delta symbol. All simulations were performed in the disordered phase at the infinite volume transition point $\beta_t = \ln(1+\sqrt{q})$ for the cases $q = 10, 15$, and $20$. As in our previous studies [10] we minimized autocorrelation times by updating the spins with the single-cluster algorithm. To take advantage of translational invariance we always employed periodic boundary conditions. By using very large lattices of size $V = L \times L$ with $L = 300, 120$, and $80$ for $q = 10, 15$, and $20$, respectively, we achieved that, starting from a disordered config-

*WJ thanks the DFG for a Heisenberg fellowship and SK gratefully acknowledges a fellowship by the Graduiertenkolleg "Physik and Chemie supramolekularer Systeme". Work supported by computer grants HLRZ hkf001 and NVV bvpf03.

uration, no tunnelings into the ordered phase occurred during the simulation. The code was implemented on a T3D parallel computer in a trivial way by running 64 time histories in parallel. In this way we generated a very high statistics of $600\,000\,\tau_{\text{int},e}$ sweeps for $q = 10$, $9\,000\,000\,\tau_{\text{int},e}$ for $q = 15$, and $4\,200\,000\,\tau_{\text{int},e}$ for $q = 20$, where $\tau_{\text{int},e}$ ($\approx 59$, 18, and 25) is the integrated autocorrelation time of the energy. The error bars are estimated by jack-kniving over this statistically independent sample.

In this work we mainly focussed on the cluster-diameter distribution $G^{\text{diam}}(x)$ which is defined [8] as the probability

$$G^{\text{diam}}(x) = \mu(\,\text{diam}\,C_{i_0} = x) \qquad (2)$$

that the cluster $C_{i_0}$ connected to a lattice site $i_0$ has a given diameter $x$, where the cluster diameter $\text{diam}\,C_{i_0}$ is the maximal extension in any of the $D$ coordinate directions; see Fig. 1. Since we employ periodic boundary conditions, we have averaged $G^{\text{diam}}(x)$ over all lattice sites $i_0$. In practice this amounts to histograming the cluster diameter weighted by the cluster size. The theoretically expected asymptotic behaviour of $G^{\text{diam}}(x)$ is an exponential decay governed by the correlation length $\xi_d$ [8],

$$G^{\text{diam}}(x) = a \exp(-x/\xi_d). \qquad (3)$$

Taking the logarithm of $G^{\text{diam}}$ and performing linear two-parameter fits it is then straightforward to extract $\xi_d$.

For comparison we considered also in the new simulations the $k_y^{(n)} = 2\pi n/L$ momentum projections ($i = (i_x, i_y)$),

$$g^{(n)}(i_x, j_x) = \frac{1}{L} \sum_{i_y, j_y} G(i,j) e^{i k_y^{(n)} (i_y - j_y)}, \qquad (4)$$

of the spin-spin correlation function

$$G(i,j) \equiv \langle \delta_{s_i s_j} - \frac{1}{q} \rangle = \frac{q-1}{q} \langle \Theta(i,j) \rangle, \qquad (5)$$

where the last equality gives the improved cluster estimator with $\Theta(i,j) = 1$, if $i$ and $j$ belong to the same cluster, and $\Theta = 0$ otherwise. As

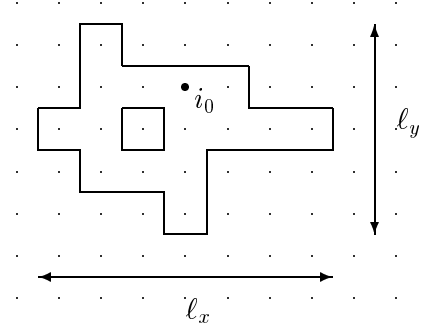

Figure 1. Illustration of the definition of the cluster diameter $\text{diam}\,C_{i_0} = \max\{\ell_x, \ell_y\}$.

discussed previously [10], here non-linear four-parameter fits of the form

$$g^{(0)}(x) = a\,\text{ch}(\frac{L/2 - x}{\xi_d}) + b\,\text{ch}(c\frac{L/2 - x}{\xi_d}) \qquad (6)$$

are necessary to extract $\xi_d$.

## 3. RESULTS

In all simulations we monitored the energy and magnetization to convince ourselves that the system never tunneled into the ordered phase. As a more quantitative measure we also recorded the maximal cluster diameter and computed energy and magnetization moments which can be compared with exact or series expansion results. As a result we are convinced that, despite the very long run times, our results for $\xi_d$ can be identified with the pure phase correlation length.

The data for $G^{\text{diam}}(x)$ and $g^{(0)}(x)$ are shown for $q = 10$ in the semi-log plot of Fig. 2. The continuous lines are one- and three-parameter fits to the Ansatz (3) and (6), respectively, with $\xi_d$ held fixed at its theoretical value $10.5595\ldots$. We see that this yields an excellent description of the fall-off of $g^{(0)}$ over more than four decades. Still, from a four-parameter fit over the same $x$ range with $\xi_d$ as a free parameter we estimate an about 10% smaller value of $\xi_d = 9.5(4)$. This confirms our earlier observation that four-parameter fits to $g^{(0)}$ systematically underestimate $\xi_d$. Compared with Ref. [10] the statistics of the present simulations is higher by more than one order of magnitude. This allowed us to include larger $x$ values in the fits and, as expected, improved the estimates of $\xi_d$, in particular for $q = 15$ and 20. An attempt to reduce the remaining discrepancies by further increasing the statistics, however, would be a rather expensive enterprise.

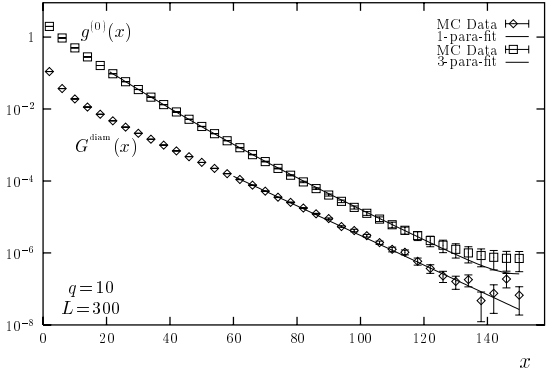

Figure 2. Semi-log plot of the cluster-diameter distribution $G^{\rm diam}(x)$ and the projected correlation function $g^{(0)}$ for $q = 10$ at $\beta_t$ in the disordered phase.

Let us therefore consider the cluster-diameter distribution which turned out to be a much cheaper solution of this problem. At first sight the constrained one-parameter fit to $G^{\rm diam}$ in Fig. 2 looks less perfect, since the data points are randomly scattered around the fit. The reason is that the correlations between the estimates at $x$ and $x + \Delta x$ are much smaller for $G^{\rm diam}$ than for $g^{(0)}$. This can be understood by noting that a cluster of diameter $x_0$ contributes only to the one estimate of $G^{\rm diam}(x)$ at $x = x_0$, but to all estimates of $g^{(0)}(x)$ with $x \leq x_0$ (recall the cluster estimator (5)). The correlation lengths resulting from unconstrained fits to $G^{\rm diam}$ are collected in Table 1. We see that the results are now in very good agreement with the theoretically expected values, with slight systematic deviations of only about $1\% - 2\%$. Contrary to $g^{(0)}(x)$ the fitted values are now overestimates. This tendency becomes obvious in Fig. 3 where we show the effective correlation lengths $\xi_d^{\rm eff} = 1/\ln[C(x)/C(x+1)]$, with $C = G^{\rm diam}$ or $g^{(0)}$. Recall that neighboring values of $G^{\rm diam}$ are much less correlated than those of $g^{(0)}$. This explains the much larger error bars on the data from $G^{\rm diam}$. For $q = 15$ and 20 we observed a similar behaviour but, as is demonstrated in Fig. 3 for $q = 20$, here the $\xi_d^{\rm eff}$ of $G^{\rm diam}$ develop a much more pronounced plateau before the statistical errors increase and the data start to fluctuate around the theoretically expected value.

Further investigations of the Ising and 3-state Potts model revealed, however, that the new observable is not always advantageous. Our results for the Ising model in the disordered phase at $\beta = 0.70340888 \approx 0.8\beta_c$ are shown in Fig. 4. Here the $\xi_d^{\rm eff}$ derived from $G^{\rm diam}$ clearly overshoot the exact value of $\xi_d = 2.6202906\ldots$ before they slowly approach it from above. The $\xi_d^{\rm eff}$ of $g^{(0)}$, on the other hand, coincide with the exact value already for very small $x$, and a simple two-parameter fit of the form (6) with $b = c = 0$ in the range $x = 1 \ldots 40$ yields $\xi_d = 2.62029(14)$, in excellent agreement with the exact value. The results for the 3-state model look very similar.

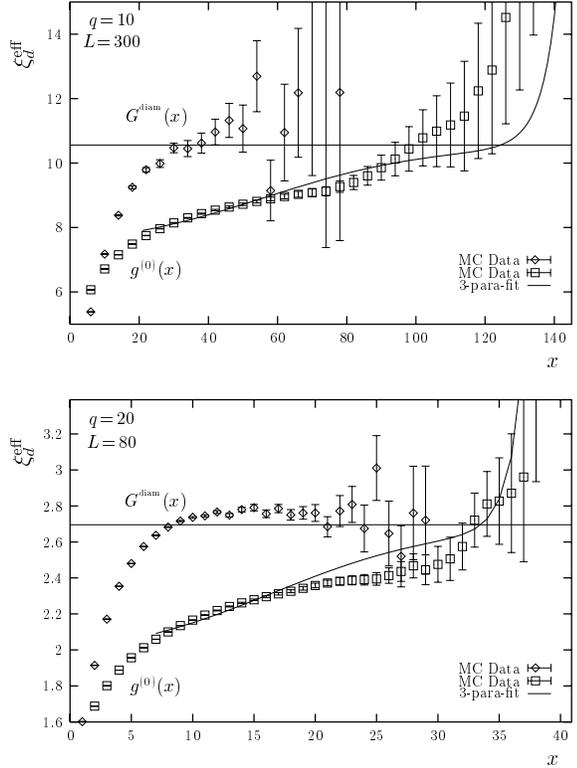

Figure 3. Effective correlation lengths for $q = 10$ and 20 at $\beta_t$ in the disordered phase derived from the correlation functions as shown in Fig. 2.

Table 1
Correlation length $\xi_d(\beta_t)$ in the disordered phase extracted from the cluster-diameter distribution

| $q = 10$ | | $q = 15$ | | $q = 20$ | |
| $300 \times 300$ | | $120 \times 120$ | | $80 \times 80$ | |
| $x_{\min} - x_{\max}$ | $\xi_d$ | $x_{\min} - x_{\max}$ | $\xi_d$ | $x_{\min} - x_{\max}$ | $\xi_d$ |
|---|---|---|---|---|---|
| 40–130 | 10.90(2) | 20–50 | 4.297(4) | 13–40 | 2.766(3) |
| 64–130 | 10.84(5) | 29–50 | 4.26(2) | 19–40 | 2.744(7) |
| 88–130 | 10.5(2) | 38–50 | 4.25(4) | 25–40 | 2.70(3) |
| exact | 10.559519... | | 4.180954... | | 2.695502... |

Also for the three-dimensional $q$-state Potts model, which undergoes a first-order phase transition already for $q \geq 3$, plots of the effective correlation lengths derived from $G^{\text{diam}}$ and $g^{(0)}$ look qualitatively as in Fig. 4. Here we simulated the models with $q = 3$, 4, and 5 at the transition point $\beta_t$ in the disordered phase. Also in these cases we found that $g^{(0)}$ gives much more reliable estimates of $\xi_d$.

## 4. DISCUSSION

Our numerical results clearly show that the cluster-diameter distribution is very well suited to extract the correlation length $\xi_d(\beta_t)$ of two-dimensional $q$-state Potts models with relatively large values of $q$. While analyses of the standard (projected) two-point function are plagued by large systematic errors, with the new observable we succeeded to reproduce the theoretically expected values at a $1\% - 2\%$ level.

For small values of $q$ and also for the three-dimensional model at the first-order transition point, however, the standard correlation function gives much more reliable results. For reasons not well understood to date, the two quite different correlators seem to behave complementary to each other.

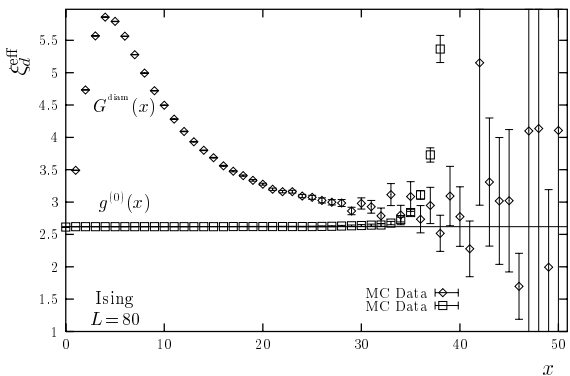

Figure 4. Effective correlation lengths for the Ising model at $\beta = 0.70340888$.